\documentstyle[12pt,sprocl,epsf]{article}
\catcode`\@=11
\long\def\@makefntext#1{
\protect\noindent \hbox to 3.2pt {\hskip-.9pt
$^{{\ninerm\@thefnmark}}$\hfil}#1\hfill}		

\def\@makefnmark{\hbox to 0pt{$^{\@thefnmark}$\hss}}  

\def\ps@myheadings{\let\@mkboth\@gobbletwo
\def\@oddhead{\hbox{}
\rightmark\hfil\ninerm\thepage}
\def\@oddfoot{}\def\@evenhead{\ninerm\thepage\hfil
\leftmark\hbox{}}\def\@evenfoot{}
\def\sectionmark##1{}\def\subsectionmark##1{}}
\def\ov{\over\displaystyle\strut}
\def\dst{\displaystyle\strut}
\def\ben{\begin{eqnarray}}
\def\enn{\end{eqnarray}}

\begin{document}

\title{HOW TO DISTINGUISH HYDRODYNAMIC MODELS
        UTILIZING PARTICLE CORRELATIONS AND SPECTRA ?}
\author{T. CS\"ORG\H O}
\address{Department of Physics, Columbia University,
538 West 120th, New York, NY 10027\\
MTA KFKI RMKI, H -- 1525 Budapest 114,
POB 49, Hungary\\
E-mail: csorgo@sunserv.kfki.hu}

\author{S. NICKERSON and D. KIANG}
\address{
Department of Physics, Dalhousie University, Halifax, N.S., Canada B3H 3J5\\
E-mails: snick@is.dal.ca and dkiang@fizz.phys.dal.ca
}
\maketitle \abstracts{
We demonstrate on examples that a
{\it simultaneous} study of the Bose-Einstein correlation function
and the invariant momentum distribution can be very useful in
distinguishing various hydrodynamic models, which describe
separately the short-range correlations in high energy
hadronic reactions as
measured by the NA22 collaboration.
We also analyze Bose-Einstein correlation
functions, measured by the NA44 experiment at CERN SPS,
in the context of the
core-halo model.
Values for the core radius and the fraction of direct bosons are
obtained,  and found to be {\it independent}
of the structure of the correlation function at small relative
momenta of $Q \le 40$ MeV.
}

\section{ Introduction }
The NA22 Collaboration performed recently a detailed study of Bose-Einstein
Correlation Functions (BECF-s) in one- two- and three-dimensions\cite{na22},
for $\pi^+ / K^+ + p$ reactions at 250 GeV.
Their study concluded as follows:
``Our data do not confirm the expectation from the string-type model ... .
 A good description of our data is ... achieved in the framework of the
hydrodynamic expanding source models\cite{pratt,bjorken,sinyu,1d,chap,3d,csl95}
... .
Alternatively, our data are also described in a non-expanding,
surface-emitting fireball-like sources\cite{kp1,kp2}... .''
The above cited hydrodynamic models,
successful in describing the NA22 data,
belong to the following {\it three} different classes:
{\it i)}	Non-expanding, spherically symmetric fireballs\cite{kp1,kp2}.
{\it ii)}	Expanding, spherically symmetric shells\cite{pratt,bjorken}.
{\it iii)}	Longitudinally expanding, cylindrically symmetric models
		with possible transverse flow, transverse and temporal
		temperature profiles\cite{sinyu,1d,chap,3d,csl95}.

\section{ Particle correlations and spectra in various
	hydrodynamic models }
The  experimental evidence reported in ref.~\cite{na22}
 indicates that it is rather difficult
to distinguish the Fourier-transformed emission functions
of the models of type {\it i) -- iii)}. However, the same emission
function, which determines the two-particle BECF-s,
 prescribes also the single-particle
spectra. In the forthcoming, we shall summarize the particle
spectra and the BECF-s of the models and discuss what are the
essential differences among these.
We shall utilize the Wigner-function formalism
along the lines of refs.\cite{pratt,2646,chap,3d},
where the particle emission is described by
the emission function $S(x,p)$, $x = (t,{\bf r})$ and
$p = (E,{\bf p})$ with $E = \sqrt{| {\bf p} |^2 + m^2}$.
The two-particle BECF for completely chaotic sources is
given\cite{2646,zajc,nr,chap} by
\ben
C(K,\Delta k) &  = &
	{\dst \langle n \rangle^2 \ov \langle n (n-1) \rangle}
		\, {\dst N_2( {\bf p}_1 , {\bf p}_2)   \ov
        	N_1( {\bf p_1} ) \, N_1({\bf p}_2 ) }
 \simeq   1 +  {\displaystyle\strut
         | \tilde S(\Delta k , K) |^2 \ov | \tilde S(0,K)|^2 },
\enn
where
$
 N({\bf p})  =  E dN/d{\bf p} = \tilde S(0,p)
$
 stands for the single-particle
inclusive invariant momentum distribution, normalized to the mean
multiplicity as $\int N({\bf p}) \, d{\bf p}/E \, = \, \langle n \rangle $ and
$
\tilde S(\Delta k , K )  =  \int d^4 x \,
                 S(x,K) \, \exp(i \Delta k \cdot x )$ with
$\Delta k  = p_1 - p_2$,  and
$K  = {(p_1 + p_2) / 2}$,
where $\Delta k \cdot x $ stands for the inner product
of the four-vectors.

\subsection{ Spectra and Correlations for  Kopylov and
Podgoretskii model}
The source considered by Kopylov and Podgoretskii (KP) in refs.\cite{kp1,kp2}
was one of the earliest models of particle emission in multiparticle
physics: a uniformly illuminated, surface emitting sphere.
The emission function itself was not discussed in detail,
since the source was assumed to be {\it non-expanding,
static}. If we assume that the source is thermalized,
the KP emission function is given by
\ben
	S_{KP}(x,p) & = &
	{\dst g \ov (2 \pi \hbar)^3 } \, E_c \,
	\delta ( R_{KP} - | {\bf r} | ) \,
	\exp( - t/\tau_{KP} ) \,
	{\dst 1 \ov \exp(E_c /T) - 1},
\enn
which is defined {\it in the  CMS of the source}.
The degeneracy factor is indicated by $g$, the pre-factor $E_c$ is due to
the invariant normalization of the momentum distribution.
The parameter $R_{KP}$ is the radius of the source,
$\tau_{KP} $ is the decay time of the quanta, or in another interpretation,
the thickness of the emitting layer, $E_c$ is the energy of the particles
in the CMS of the source and $T$ is the temperature of the fireball.
The single-particle spectrum can be obtained as
\ben
	E {\dst dN_{KP}\ov d{\bf p} } & = &
	N_{KP}({\bf p} ) \,\,
	  =   \,\,   {\dst g \ov (2 \pi \hbar)^3 } \,
		E_c V_{KP} \,
		{\dst 1 \ov \exp(E_c /T) - 1},\label{e:nkp}
\enn
where $ V_{KP} =  4 \pi R^2_{KP} \tau_{KP}$ is the effective volume of the
fireball.
In Boltzmann approximation,  this
invariant momentum distribution can be rewritten as
\ben
	N_{KP}({\bf p}) & =  &
			 {\dst dN \ov 2 \pi m_t dm_t dy}
			\simeq     {\dst g \ov (2 \pi \hbar)^3 }
		\,	m_t \cosh(y - y_0)V_{KP} \,
			\exp\left( - {\dst m_t \ov T_{KP}(y)} \right), \\
	T_{KP}(y) & = & {T / \cosh( y - y_0) }, \label{e:tkp}
\enn
where
$m_t$ and
	$ m_t = \sqrt{p_x^2 + p_y^2 + m^2} $ is the transverse mass,
	$y = 0.5 \log(( E + p_z)/(E-p_z)) $ denotes the rapidity
	of the particle, and $ y_0$ stands for the rapidity
	of the CMS of the fireball.
	The effective rapidity-dependent temperature distribution ,
	as given by eq.~(\ref{e:tkp}),
	decreases fast with increasing difference
	between the rapidity of the particle as compared to
	the CMS of the fireball. This $1/\cosh(y - y_0)$
decrease is a typical result for non-expanding thermalized fireballs,
as discussed in ref.\cite{schned}.
The spectrum of eq.~(\ref{e:nkp}) can be re-written approximately as
\ben
	{\dst dN \ov 2 \pi m_t dm_t dy}
		& \simeq &    {\dst g \ov (2 \pi \hbar)^3 }
		\,m_t \cosh(y - y_0) V_{KP} \,
		\exp\left( - {\dst (y - y_0)^2  \ov 2 \Delta\eta_T^2}\right),
		\label{e:dekp}
\enn
where $\Delta\eta_T^2   =   T / m_t$ .
This yields a specific transverse mass dependence for the
rapidity-width of the spectrum at a given value of $m_t$, which can be
checked easily in experimental data analysis.
The BECF for the KP emission function is given as
\ben
C(q_T, K) & = & 1 +
		\lambda \, [I(R_{KP} q_T)]^2  \,
			(1 + \tau_{KP}^2 q^2_0)^{-1},
	\label{e:ckp}
\enn
where $I(x) = 2 J_1(x)/x$
and $J_1(x)$ is the Bessel function of first order,
with $I(0) = 1$ and $I(\infty) = 0$.
The parameter $\lambda $ is introduced phenomenologically to
account for intercepts $ 1  + \lambda < 2$, $q_T =
		{\bf \Delta k} \cdot {\bf K} / |{\bf K}|$
is the relative momentum transverse to $K$ and
$q_0 = E_1 - E_2$ is the energy difference in the CMS.
{\it This BECF is independent of the mean momentum ${\bf K}$
due to the non-expanding source described by the KP model.}

\subsection{Spectra and Correlations for  Expanding Shells }
	The model introduced by S. Pratt (P) and discussed recently
	by J. Bjorken corresponds to a uniformly illuminated,
	expanding shell\cite{pratt,bjorken}, given by
\ben
        S_{P}(x,p) & = &
        {\dst g \ov (2 \pi \hbar)^3 }\, E_c \,
        \delta ( R_{P} - | {\bf r} | )
        \exp( - t^2/\tau_{P}^2 )
        {\dst 1 \ov \exp(p \cdot u(| {\bf r}|) /T) - 1},
\enn
which is defined {\it in the  CMS of the source}.
The parameter $R_{P}$ is the radius of the source,
$\tau_P $ is the decay time of the quanta or
the thickness of the emitting layer
in the CMS of the source
and $ u({\bf r}) = \gamma (1, v {\bf r} / R_P)$ is
a spherically symmetric flow profile characterizing the expansion
of the shell, with $\gamma = 1/\sqrt{1-v^2}$.
This model corresponds to a three-dimensional, spherically
symmetric expansion in the rest frame of the source, and the
KP model is recovered in the $v = 0$ limiting case.

The single-particle spectrum can be evaluated
in Boltzmann approximation as
\ben
     {\dst dN \ov 2 \pi m_t dm_t dy}
     & \simeq &    {\dst g \ov (2 \pi \hbar)^3 }
                      	\, m_t \cosh(y-y_0)V_P \,
			{\dst \sinh( a) \ov a}\,
			\exp( - m_t/T_{P}(y)) \\
	T_{P}(y) & = & {\dst T \ov \gamma\cosh( y - y_0) }
			\, [ 1 + {\cal O}(v^2 \gamma^2)] \label{e:tp}
\enn
where $ V_P =  4 \pi R^2_P (\pi \tau_P)^2$
is the volume of the fireball, $ a = a({\bf p} ) =
v \gamma | {\bf p} |/T  $ is a momentum-dependent parameter
related to $v$, the surface velocity of the shell and
$T_P(y)$ is the effective rapidity - dependent slope-parameter
distribution.
Thus for small values of the parameter
$a({\bf p})$ the effective temperature
is decreased  as compared to the static fireball (KP) case, however,
the rapidity-width of this effective temperature distribution
remains unchanged even if a small spherical expansion is included.
The BECF for the spherically expanding shells can be written
in a directional dependent,
 analytic form as given by Eq.~(13) of ref.\cite{pratt}.
Although this analytic form is too complicated to be repeated here,
one arrives at a simpler expression\cite{pratt},
\ben
 	\langle C(q, K) \rangle & = & 1 + \lambda \,
			\exp( - 0.5 \, R_{*,P}^2 \, q^2) \,
			\exp( - 0.5 \, \tau^2_P\, q_0^2) ,
			\label{e:cpb}
\enn
after averaging over the direction of the relative momentum in the CMS of
the source.
Here $q = |{\bf \Delta k } |$ in the CMS
and the radius parameter reads\cite{pratt} as
\ben
	R_{*,P} & = & R_P [( a({\bf p}) \tanh(a({\bf p}) )^{-1}
		- \sinh( a({\bf p}) )^{-2} ]^{1/2} . \label{e:rsp}
\enn
 	The momentum dependence of the effective radius parameter
$R_{*,P}$ is a direct consequence of the expansion of the source.

\subsection{Spectra and Correlations for  Longitudinally
Expanding Systems }

	Systems which are dominantly expanding longitudinally
	appear to be the relevant models to high energy heavy ion
	reactions at CERN SPS energy region, $\sqrt{s} = 20 GeV A$.
	The detailed presentation and elaboration of these type
	of models is outside of the scope of the present contribution,
	we recommend refs.\cite{1d,3d,mpd25,chap,csl95} for further
	details.
  However, we summarize here approximate results for the single-particle
	spectra:
\ben
     {\dst dN \ov 2 \pi m_t dm_t dy}
     & \simeq &    {\dst g \ov (2 \pi \hbar)^3 }
                      	\, m_t \cosh(y- \eta_s({\bf p}))V_*({\bf p}) \,
			\exp( - m_t/T_{L}(y)), \\
	 {\dst dN \ov 2 \pi m_t dm_t dy}
     & \simeq &    {\dst g \ov (2 \pi \hbar)^3 }
        \, m_t \cosh(y- \eta_s({\bf p}))V_*({\bf p}) \,
	\exp\left( - {\dst (y - y_0)^2 \ov 2 \Delta y(m_t)^2} \right), \\
	V_*({\bf p}) & \simeq & V_0 \left( T/m_t\right )^{\alpha - 1}, \\
	T_{L}(y) & = & {\dst T_* \ov 1 + a_T ( y - y_0)^2 },
	\qquad \hbox{\rm and} \qquad
 	\Delta^2 y(m_t)  =  \Delta^2\eta + T/m_t	. \label{e:ymt}
\enn
	These relations indicate that the rapidity dependence of the
	temperature parameter $T_L(y)$ can be described with
	a new fit parameter $a_T$ (which was shown\cite{3d} to be related to
	$\Delta\eta$, the
	total longitudinal extension of the particle emitting source,
	e.g. $a_T = 0$ for infinite systems).  A more direct access to
	the longitudinal size of this expanding system is provided by
	the transverse mass dependence of the rapidity-width of the
	invariant momentum distribution\cite{1d,3d}, because
	$\Delta^2 y(m_t) \approx \Delta^2\eta$ for $m_t >> T$.

	The BECF-s for such systems
	can be written to the following form:
\ben
	C({\bf \Delta k},{\bf K}) & = &
	1 +\lambda_*({\bf K}) \exp\left( - R_{side}^2({\bf K}) Q_{side}^2
			 - R_{out}^2({\bf K}) Q_{out}^2
			 - R_{long}^2({\bf K}) Q_{long}^2\right) \times
							\nonumber\\
	\null & \null & \phantom{ 1 +\lambda_*({\bf K})}
			\exp\left(
			 - R_{olong}^2({\bf K}) Q_{out} Q_{long}\right),
\enn
	where the intercept parameter can be interpreted\cite{halo}
	in the core-halo model, the side, out and longitudinal radius
	parameters as well as the out-long cross-term\cite{chap}
	 may in general depend on the mean momentum ${\bf K}$
	in a complicated manner~\cite{1d,chap,3d}. However, in some
	specific limiting cases\cite{3d} this can be simplified
	as $R_{side}\simeq R_{out} \simeq R_{long} \propto 1/\sqrt{m_t}$.

\subsection{How to distinguish the three model-classes?}
The following tests can be performed experimentally to
check the hypothesis of Kopylov and Podgoretskii, Pratt and Bjorken
or the model-class of longitudinally expanding, finite systems
 in a more detailed
manner:
\begin{enumerate}
\item{} Measure the $m_t$ dependence of the effective
	rapidity-width of the $N({\bf p})$
	distribution at a fixed value of $m_t$, and try to fit the
	result with Eq.~(\ref{e:ymt}).
	For static or spherically expanding shells, the parameters
	$\Delta y(m_t)$ decrease to 0 as $1/\sqrt{m_t}$, while
	for longitudinally expanding finite systems the large $m_t$
	limit is $\Delta y(\infty)= \Delta\eta > 0$, the longitudinal
	size of the system in space-time rapidity.
\item{} Measure the rapidity dependence of the temperature parameter
	of the  $N({\bf p})$
	distribution at a fixed value of $y$, and try to fit the
	result with Eq.~(\ref{e:ymt}). A slow drop of the
	effective temperature with increasing values of $|
	y - y_0 |$ results in the breakdown of the static
	fireball picture. Longitudinally very extended,
	expanding systems are predicted to have a Lorentzian
	effective temperature distribution, $T_{eff}(y) = T_*/
	(1 + a_T (y - y_0)^2 )$ where the parameter $a_T$ carries
	information about the longitudinal extension of the
	source\cite{3d}.
\item{} Check if the parameters $R_{KP}, R_P, \tau_P$ and
	$\tau_{KP}$ of the Bose-Einstein correlation
	function of eq.~(\ref{e:ckp}) are independent or not
	of various values of
	 ${\bf K}$.
	Static models predict
	$R_{KP}({\bf K}) = const$ and $\tau_{KP}({\bf K}) = const$,
	a deviation from this behavior results in the breakdown
	of the KP  or any other static model.
	Spherically symmetric expanding shells also
	have a specific transverse momentum dependent $R_{*,P}({\bf K})$
	parameter, decreasing with increasing values of $|{\bf K} |$
	as given by eq.~(\ref{e:rsp}). Longitudinally expanding
	finite systems have  momentum dependent longitudinal
	radius parameter, the transverse radius parameters may or may not
	be transverse mass dependent\cite{1d,3d}.
\end{enumerate}
 	Let us now focus on the
	interpretation of the intercept parameter of the
	BECF-s.
\section{ Core-halo model analysis of NA44 data  }
  Having first performed a Monte Carlo simulation to justify the
analysis technique to be used, we then analyze the Bose-Einstein correlation
functions from NA44\cite{na44,sk}.
\eject

\subsection{Monte-Carlo test of the method}
In a Monte-Carlo study, we simulate the actual
and background distribution ($A(Q)$ and $B(Q)$)
 of particle pairs.  For the CERN experiment NA44, studying
S+Pb collisions, the shape of $B(Q)$ is given approximately by
$
B(Q)=Q^3 e^{- (3.6 \, Q^{0.3})},
$
which form reproduces the experimentally measured
background distribution\cite{bengtpr}.
This background distribution  is peaked at 25 MeV similarly to the NA44 data.
We have started with a core-halo model correlation function\cite{halo},
assuming  Gaussian source functions for the core and for the halo,
choosing the core radius parameter to be $R_c = 4$ fm. The
radius parameter of the halo was taken to be $R_h = 40 $ fm
and the core fraction was $f_c = 0.7$.
The actual distribution $A(Q) $ was sampled according to the
formula
\ben
	A(Q) & = & B(Q) [ 1 +
		 f_c^2 \, \exp(-R_c^2 Q^2) + \nonumber \\
	\null &  \null &
		2 \, f_c \,(1-f_c) \, \exp( - 0.5 (R_c^2 + R_h^2) Q^2) +
		 (  1 - f_c)^2 \, \exp(-R_h^2 Q^2)
		 ]
\enn
which corresponds to the full correlation function including
correlations of $(c,c)$, $(c,h)$ and $(h,h)$ type of particle pairs.
Here $c$ refers to the core, which is assumed to be resolvable
since
$R_c < \hbar / Q_{bin}$, where the two-particle relative momentum
resolution $Q_{bin}$ is about the size of a bin (cca. 10 MeV for NA44).
In the core-halo model\cite{halo},
the halo (index $h$)  is assumed to change on large
length-scales, which are un-resolvable by the two-particle correlation
measurements, $R_h > \hbar/Q_{min}$.

Having sampled the actual and background
distributions of particle pairs in a Monte-Carlo simulation,
 the correlation function is now calculated in each bin of size $Q_{bin}$
as
$ C_2(Q)=\frac{A(Q)}{B(Q)}.$
This form is fitted by the expression
\ben
C_2(Q) = 1 + \lambda_* \, \exp(-R_*^2 Q^2) .
\enn

If the  core-halo model is applicable,
it predicts that $\lambda_* = f_c^2$ and
$R_* = R_c$ after the fitting,
i.e. the intercept can be utilized to measure the
fraction of particles emitted by the core, and $R_*$ the radius parameter
will coincide with the radius parameter characterizing the core.
Furthermore, if the assumptions of the core-halo model are satisfied
by some data, then the fitted $\lambda_*$ and $R_*$
parameters should become independent of the fitted region, obtained
by excluding more and more part of the correlation function at
the lowest values of $Q$, as long as the inequalities
$\hbar/R_h < Q_{min} < \hbar/R_c$ are satisfied by $Q_{min}$.
We chose the range from $Q_{bin} = 10 \le  Q_{min} \le 60 $ MeV.

We have checked in the Monte-Carlo simulation
if the effect of the halo on
the  deviation from a Gaussian form does really cancel or not
from the simulation by increasing the
value of $Q_{min}$, the size of the low $Q$ region excluded from the fit.
We have found that indeed $\lambda_* = f_c^2\simeq 0.5$ and $R_* = R_c = 4.0
$ fm was reproduced by the fitting within the statistical uncertainties
even for as large values of the excluded region as $Q_{min} = 60 $ MeV,
when already more then one half of the peak was removed from the
correlation function, utilizing 300 K pairs.
Armed with this experience, we then applied
the method to the analysis of NA44 data for pions and kaons in
various one-dimensional slices of the correlation functions at  given
 values of ${\bf K}$.

\subsection{ Analysis of NA44 data}

In order to extract values for $R$ and $f_c$ from data, we
first fitted the side, out and longitudinal slices of the
NA44 data\cite{na44}
on $S + Pb$ reactions at 200 AGeV for pions at low and
high transverse mass as well as for kaons.
Thus we obtained radius parameters and intercept parameters
(with their errors, respectively), as a function of $Q_{min}$.
We determined these fit parameters in the $Q_{min} = 0 - 60 $
MeV region.
The fitted $R_*(Q_{min})$ and $\lambda_*(Q_{min})$
parameters up to $Q_{min} = 40 $ MeV/c were (meta)-fitted	with a
$Q_{min}$ independent constant.  Our findings are summarized
in Table 1. More details of the study
shall be reported elsewhere\cite{sk}.

\begin{table}[h]
\centering
\begin{tabular}{|c|c|c|c|}
\hline\hline
parameter & high $p_t$ $\pi^+$ & low $p_t$ $\pi^+$ & $K^+$ \\ \hline
$R_{c \ out}$ (fm)& 2.92 $\pm$ .13 & 4.29 $\pm$ .13 & 2.54 $\pm$ .18 \\
$R_{c \ side}$ (fm)& 2.90 $\pm$ .18 & 4.24 $\pm$ .26 & 2.22 $\pm$ .19 \\
$R_{c \ long}$ (fm)& 3.31 $\pm$ .16 & 5.43 $\pm$ .30 & 2.67 $\pm$ .22 \\
$f_{c \ out}$ & .704 $\pm$ .012 & .725 $\pm$ .011 & .802 $\pm$ .027 \\
$f_{c \ side}$ & .735 $\pm$ .021 & .647 $\pm$ .021 & .736 $\pm$ .033 \\
$f_{c \ long}$ & .738 $\pm$ .014 & .724 $\pm$ .015 & .789 $\pm$ .029 \\
\hline\hline
\end{tabular}
\caption{\small Extracted values for $R_c$ and $f_c$ from NA44 S+Pb data.}
\label{results}
\end{table}
In ref.\cite{halo} the halo has been interpreted as being
created essentially
by the decay products of the $\omega$, $\eta$ and $\eta^{\prime}$ resonances.
The measured core fractions of Table 1 are indeed in
 the vicinity of the core fractions of Fritiof and RQMD\cite{restable}
 if $\omega$, $\eta$ and $\eta^{\prime}$ are taken
as unresolved long lived resonances, as indicated by Table ~\ref{comp}.
\begin{table}[h]
\centering
\begin{tabular}{|c|c|c|c|}
\hline\hline
$f_{c,NA44}$ (high $p_t$ $\pi^+$ ) & $f_{c,NA44}$ (low $p_t$ $\pi^+$ )
& $f_{c,Fritiof}$ & $f_{c,RQMD}$ \\ \hline
        0.71 $\pm$ 0.01 & 0.71 $\pm$ 0.01 & 0.75        &       0.80 \\
\hline\hline
\end{tabular}
\caption{\small Core fractions $f_c$ from NA44 S+Pb data$^{19}$
compared to core fractions from RQMD and Fritiof,
when $\omega$, $\eta$ and $\eta^{\prime}$ are taken as unresolved
long-lived resonances$^{17}$.
The core fractions for the NA44 low $p_t$ and
high $p_t$ sample are obtained from a simultaneous fit to
($f_{c,out},f_{c,side},f_{c,long}$) by a constant value, and the result
is rounded to two decimal digits. }
\label{comp}
\end{table}
Thus we do {\it not}
find a complicated, non-Gaussian structure of the measured
BECF in $S + Pb$ reactions at CERN SPS, in contrast to some
earlier expectations which argued that resonance decays,
coupled to a freeze-out hyper-surface obtained from
(hydro)dynamic evolution may lead to resolvable deviations
from a Gaussian structure.
Such a deviation has been predicted by detailed simulations of resonance
decay effects by calculations with SPACER\cite{2646},
HYLANDER\cite{res1}, and recently by H. Heiselberg\cite{restable}.
Assuming that such non-Gaussian structures
{\it are} created by the effect of $\eta, \eta^{\prime}$ and $\omega$ resonance
decays to pions, but that they are experimentally {\it not yet
resolvable} in the NA44 $S + Pb \rightarrow 2 \pi + X $
experiment,  we obtain a consistent interpretation
of the data and a physical interpretation of the
measured intercept and radius parameters\cite{halo,sk,3d}.

\section*{\bf Acknowledgments:}
Cs. T. would like to thank D. Kiang for kind hospitality at the
Dalhousie University, Halifax, and W. Kittel
for  invitation to  and support at this exciting Nijmegen
workshop.  This work was supported by the
OTKA grant No.T016206,
by the U.S. - Hungarian Science and Technology
Joint Fund, No. 378/93 and by an Advanced Research Award form the Fulbright
Foundation, grant No. 20925/1996.
 We thank the Institure for Nuclear Theory at the
 University of Washington for its hospitality and the
 Department of Energy for partial support during the
 completion of this work.

\vfill\eject
\end{document}